\documentclass[%
twocolumn,
 longbibliography,
 amsmath,amssymb,
 aps,
 pre,
]{revtex4}

\usepackage{mathtools}
\usepackage{lipsum}
\usepackage{graphicx}
\usepackage{dcolumn}
\usepackage{bm}
\usepackage{comment}
\usepackage{afterpage}
\usepackage{xcolor}
\definecolor{sapphire}{HTML}{0067A5}
\usepackage[colorlinks=true,citecolor=sapphire,linkcolor=sapphire,urlcolor=sapphire]{hyperref}

\begin{document}
\title{Spontaneous crumpling of active spherical shells} 
\author{M. C. Gandikota$^1$, Shibananda Das$^{1,2}$ and A. Cacciuto$^1$}
\email{ac2822@columbia.edu}
\affiliation{$^1$ Department of Chemistry, Columbia University\\ 3000 Broadway, New York, NY 10027\\
$^2$ Department of Polymer Science and Engineering, University of Massachusetts, Amherst, Massachusetts 01003}

\begin{abstract}
The existence of a crumpled phase for self-avoiding elastic surfaces was postulated more than three decades ago using simple Flory-like scaling arguments. Despite much effort, its stability in a microscopic environment has been the subject of much debate. In this paper we show how a crumpled phase develops reliably and consistently upon subjecting  a thin spherical shell to active fluctuations.
We find a master curve describing how the relative volume of a shell changes with the strength of the active forces, that applies for every shell independent of size and elastic constants. Furthermore, 
we extract a general expression for the onset active force beyond which a shell begins to crumple. Finally, we calculate how the size exponent varies along the crumpling curve.
\end{abstract}

\maketitle
\section{Introduction}
It is well known that the energy cost required to deform an elastic surface is well accounted for by a bending and a stretching energy term~\cite{LandauBook,Love2011Jun,NelsonBook}. Despite its apparent simplicity, the coupling between bending and stretching modes of deformation is highly nonlinear for thin elastic materials such as sheets and shells giving rise to mechanical behavior that is hard to predict. 
What makes these materials particularly exciting is that the ratio between stretching, $E_s$, and bending, $E_b$, energies for an arbitrary deformation of amplitude $h$ on a surface of thickness $t$ scales as $E_s/E_b \simeq (h/t)^2$~\cite{LandauBook}. Therefore, for sufficiently thin surfaces, ($t\ll h$), only stretch-free deformations are allowed. Skin wrinkling under applied stress~\cite{Cerda2002Oct,Efimenko2005Apr}, stress focusing via cone formation of crumpled paper~\cite{Witten2007Apr}, and buckling of  thin shells~\cite{VonKarman}, are just a few examples arising from this global constraint.

The theory of elasticity developed for continuum mechanics has been successfully used to study a number of microscopic systems, including viral capsids~\cite{lidmar2003}, graphite-oxide~\cite{Spector1994Nov,wen1992crumpled} and graphene sheets~\cite{Stankovich2006Jul,Meyer2007Mar}, cross polymerized  membranes~\cite{fendler_polymerized_1984} and gels~\cite{Georges2005Apr}, the spectrin-actin network forming the cytoskeleton of red blood cells~\cite{schmidt1993existence,Lux2016Jan} and close-packed nanoparticle arrays~\cite{Mueggenburg2007Sep}. 
Significantly, at this length-scale, the nonlinear coupling between the different elastic modes can also be induced by  thermal fluctuations with significant consequences for the structure of microscopic surfaces. Thermal fluctuations renormalize the bending rigidity of thin elastic sheets which become stiffer as their size increases~\cite{peliti1987,aronovitz1988,chaikin1995} leading to the stabilization of a flat phase for two-dimensional unsupported surfaces. 
In thin spherical shells, thermal fluctuations act as an effective negative internal pressure capable of buckling the shell~\cite{kovsmrlj2017}.  

Simple Flory-like arguments~\cite{wiese2000} that work so well in establishing the scaling laws of self-avoiding polymers~\cite{Gennes1979Nov},  predict that self-avoiding  elastic sheets should be found in a crumpled state for negligible bending rigidities. This state is characterized by the scaling of the radius of gyration of the sheet, $R_{\rm{g}}$, with its side length, $L$, of the form $R_{\rm g}\sim L^{\nu}$ with the Flory exponent $\nu=4/5$~\cite{domb2001,paczuski1988}. Yet, numerical simulations of fully crystalline, self-avoiding elastic sheets indicate that these surfaces always acquire a rough, but overall extended (flat), state with a size exponent along their longitudinal directions $\nu\simeq 1$, even in the absence of bending rigidity~\cite{plischke1988,abraham1991folding,bowick2001,abraham1989}.
As of today, it is fair to say that the existence of a crumpled phase for large tethered membranes in equilibrium remains uncertain. 

Inspired by early experiments of graphite-oxide sheets in poor solvent~\cite{Wen1992Jan} which seemed to indicate the presence of a crumpled phase upon improving the quality of the solvent, simulations were performed to include
the presence of attractive interactions, but the crumpled phase was not observed~\cite{abraham1989}.
Interestingly, thin elastic spherical shells in the presence of explicit attractive forces were initially reported to have a Flory exponent compatible with a crumpled phase in a temperature window that falls between the flat and the compact regimes~\cite{liu1992}, but simulations with larger shells did not find such intermediate regime~\cite{grest1994}.
For a comprehensive review on the subject, we refer the reader to references~\cite{NelsonBook,wiese2000,bowick2001,nelson2004,kovsmrlj2017}. 
What is certain at this point is that the one way to reliably obtain a crumpled phase out of elastic thin surfaces is by quickly compressing them  using a large external force~\cite{kantor1988,gomes1987,kantor1987,gomes1989}, or by rapidly dehydrating graphene-oxide nanopaper~\cite{ma2012}. 

In this paper, we reconsider this problem within the framework of active matter, and study the effect of active, non-equilibrium fluctuations on the structure of a thin spherical shell. Crucially, we show how a crumpled phase, develops systematically and reliably  for sufficiently large active forces. While a significant amount of work has been done to understand the structural and dynamic properties of active linear  and  ring polymers~\cite{Winkler2020Jul,deviations2019,CacciutoDas2021,Gandikota2022Mar,Winkler2017Aug,mousavi2019active}, and more recent work considered the behavior of fluid vesicles in the presence of active fluctuations or active agents~\cite{Iyer2022Sep,Cagnetta2022Jan,Turlier2018Dec,Kulkarni2023Apr,Iyer2023Jan}, apart from a few recent papers~\cite{gandikota2023,Mallory2015Jul,agrawal2023active}, very little is known about how elastic surfaces respond to non-equilibrium fluctuations, and this is an important problem given their relevance to biological and synthetic materials~\cite{turlier2016,Bowick2022Feb,Kusters2019Sep}. 

\section{Numerical model}
We model the thin spherical shell using a triangulated network of harmonic bonds~\cite{kantor_crumpling_1987} organized as an icosadeltahedron~\cite{Siber2020Apr}. All vertices in icosadeltahedra are six-coordinated apart from twelve five-coordinated vertices (disclinations) as required by topological constraints. This structures are characterized by two integers $(a,b)$ which define the relative position on the spherical lattice of the disclinations, and set the total number of vertices to $N=10\,(a^2+ab+b^2)+2$.
Although most of our data are obtained with this crystalline structure,  we also considered amorphous shells with a disordered distribution of nodes on the sphere. 
We construct amorphous shells by running Monte Carlo simulations of a fluid membrane (bond flip moves allows for neighbor exchange~\cite{gompper1996random}) for sufficiently long time until the bonds are randomized. We then freeze the bonds and use this as the initial configuration for our simulations. 
The rest shape of the crystalline shell is an icosahedron~\cite{lidmar2003} while that for an amorphous shell is a sphere~\cite{paulose2012}. 
 
To enforce self-avoiding interactions, we place a spherical particle of diameter $\sigma$ at each vertex. Each of these vertex particles are connected to their nearest neighbors with a harmonic potential. The interaction potential of the system can be written as
\begin{equation}\label{Hamiltonian}
\begin{split}
 U&=K\sum_{<ij>}(r_{ij}-\sigma)^2+\kappa\sum_{<lm>}(1-\bm{\eta}_l\cdot\bm{\eta}_m)\\
&+ 4\,\varepsilon\sum_{ij}\left[ \left( \frac{\sigma}{r_{ij}}\right)^{12} - \left(\frac{\sigma}{r_{ij}}\right)^{6} +\frac{1}{4}\right ],
\end{split}
\end{equation}
where $r_{ij}$ is the distance between any two vertices $i,j$.
The first term accounts for the harmonic bonds between
nearest neighbor particles with a spring constant $K$. The second term is the bending energy where, $\kappa$ is the bending rigidity of the shell and ($\bm{\eta}_l,\bm{\eta}_m$) are the normal vectors of any two adjacent triangles, $l$ and $m$, sharing an edge. 
The third term implements the excluded volume interactions between the vertex particles. The potential is cut off at $2^{1/6}\sigma$ and set to zero beyond that distance. Unless otherwise specified, we set $K=160\,k_{\rm B}T_0/\sigma^2$, $\kappa=10\,k_{\rm B}T_0$, and $\varepsilon=k_{\rm B}T_0$, where $k_{\rm B}$ is the Boltzmann constant and $T_0$ is the reference temperature.
Activity is introduced in the system by adding a self-propelling velocity of constant magnitude $v_p$ to each of the node particles. The system dynamics is resolved using Brownian motion according to~\cite{bechinger_active_2016,zottl_emergent_2016,shaebani2020computational}
\begin{equation}\label{langevin}
\begin{split}
\frac{d\pmb{r}_i(t)}{dt}  &=  \frac{1}{\gamma} \pmb{f}_i +   v_p \,  \pmb{\hat{n}}_i(t)\,  + \sqrt{2D}\,\pmb{\xi}(t),\\
\frac{d \pmb{{\hat{n}}}_i(t) }{dt}&=\sqrt{2D_r}\, \pmb{\xi}_r(t) \times \pmb{\hat{n}}_i(t),
\end{split}
\end{equation}
where  $i$ is the particle index and the unit vector $\pmb{\hat{n}}$ is the axis of propulsion. The conservative forces on each particle are denoted by $\pmb{f}_i=-\partial U/\partial \bm{r}_i$. 
 The translational diffusion coefficient $D$, temperature $T_0$ and the translational friction $\gamma$ are constrained to follow the Stokes-Einstein relation $D=k_{\rm B}T_0\gamma^{-1}$. Likewise, the rotational diffusion coefficient is constrained to be $D_r=k_{\rm B}T_0\gamma_r^{-1}$, with $D_r = 3D\sigma^{-2}$. The Gaussian white-noise terms induced by the solvent for the translational $\pmb{\xi}$ and rotational $\pmb{\xi}_r$ motions are characterized by the relations $\langle \pmb{\xi}(t)\rangle = 0$ and $\langle \xi_p(t) \xi_q(t^\prime)\rangle = \delta_{pq}\delta(t-t^\prime)$ (here the indices $p$ and $q$ stand for the Cartesian components $x,y,z$).

 The  simulations have been carried out using the numerical package LAMMPS~\cite{plimpton_fast_1995} and the units of length, time  and energy respectively are set to be $\sigma$, $\tau=\sigma^2D^{-1}$ and $k_{\rm B}T_0$. Consequently, the spring constant $K$, bending constant $\kappa$ and self-propulsion speed $v_p$ are measured in units of $k_{\rm B}T_0/\sigma^2$, $k_{\rm B}T_0$ and $\sigma/\tau$ respectively. Physically, $\tau$ is the time taken for a passive particle to diffuse a length of $\sigma$. All simulations were run with a time step smaller than $\Delta t=2\times10^{-6}\,\tau$. We record the state of the system every $10^5$ time steps. After the system reaches a steady state as indicated by the saturating values of the shell volume, we collect statistics for a time period~$\sim (10^3-10^4)\,\tau$, which amounts to a minimum of one billion time steps for the smaller systems and up to 10 billion time steps for the largest. For both crystalline and amorphous shells, most of the data we report are for elastic constants of $K=160$ and $\kappa=10$.

\section{Crumpling of active shells}

\begin{figure}[t]
	\centering
	\includegraphics[width=0.42\textwidth]{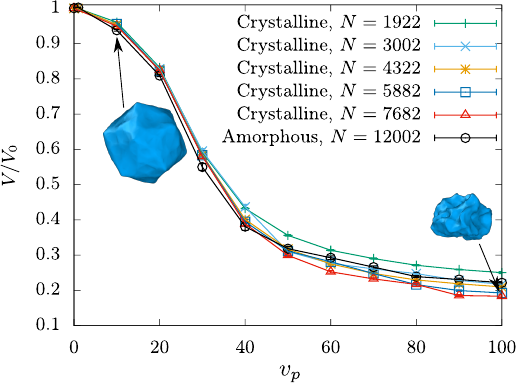}
	\caption{The monotonically decreasing normalized volume ($V/V_0$) as a function of the self-propulsion speed $v_p$, for shells of different size, $N$ and different structure (Crystalline/Amorphous). Simulation snapshots for the crystalline shell with $N=7682$ are shown at $v_p=10\text{ and }100$. The elastic constants are $K=160$ and $\kappa=10$.}\label{finite}
	\end{figure}
With our choice of elastic constants $K$ and $\kappa$, crystalline shells acquire a well-defined icosahedral state when thermalized in the absence of active fluctuations. This is expected~\cite{lidmar2003} given that with our parameters, the F$\ddot{\rm o}$ppl-von K$\acute{\rm a}$rm$\acute{\rm a}$n number $\gamma_F=\frac{4}{3}KR^2/\kappa\gg 10^2$ for every shell radius $R$ considered in this study for $N\in[1922,12002]$. 
We begin our analysis by monitoring the volume of the shell as a function of increasing values of the strength of the active forces. For small values of $v_p$, the shells develop smooth surface undulations  on the scale of the shell radius without altering its overall shape. Upon increasing the activity, these undulations deepen resulting in a monotonically decreasing shell volume and a disruption of its global symmetry. For even larger self-propulsion velocities, the shell crumples to a volume
that saturates at roughly 20\% of its original value $V_0=(4/3)\pi R^3$. See Fig.~\ref{cross-section} of Appendix for the cross-sections of locally flat and crumpled spheres.

Interestingly, repeating the same calculations with different crystalline~\cite{lidmar2003} and amorphous shell~\cite{paulose2012} sizes $N\in[1922,12002]$ result in normalized volume $V/V_0$ curves with only a weak size dependence at larger self-propulsion speeds (see Appendix~\ref{norm_vol}). Figure~\ref{finite} shows the results of this analysis. 
We find the flex of this curve, calculated by evaluating the maximum of $|dV/dv_p|$, to be a reasonable estimator of the onset self-propulsion speed, $v_p^*\approx30$,  beyond which the shells begin to crumple.  
Our results clearly show that the activity-induced crumpling of the shell is quite general and does not depend on the initial specific structure of the shell (crystalline or amorphous).
We should also stress that the maximum of  $|dV/dv_p|$ does not show a diverging peak with system size, suggesting that for this system, crumpling is a smooth process devoid of singularities rather than a real phase transition.

\begin{figure}[t]
	\centering
	\includegraphics[width=0.40\textwidth]{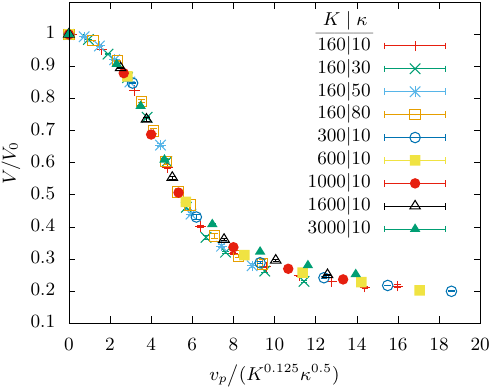}
	\caption{The collapse of normalized volumes ($V/V_0$) as a function of $v_p/(K^{0.125}\kappa^{0.5})$ for various elastic spring constants $K$ and bending constants $\kappa$. The size of the crystalline shell is $N=4322$.}\label{elastic_collapse}
\end{figure}

To understand how the onset value, $v_p^*$, depends on the elastic parameters of the membrane, we performed the same calculations with different sets of stretching and bending rigidities, in the range of  $K\in{[160,3000]}$ and $\kappa\in{[10,80]}$.
Remarkably, as shown in Fig.~\ref{elastic_collapse}, all data collapse into an empirical master curve of $V/V_0$ against $v_p/(K^{0.125}\kappa^{0.5})$. This suggests a universal crumpling onset for any shell size and any set of elastic constants beyond $v_p^*\simeq 5\,K^{1/8}\kappa^{1/2}$. 
To ensure that the collapse of  data is not an artifact of the choice of bending energy chosen in Eq. \ref{Hamiltonian}, we also considered a bending energy based on calculating squares of local averages of mean curvature~\cite{gompper1996random,guckenberger2017theory}. We find that the results stand independently of the specific discretization of the bending energy (see Appendix~\ref{discretization}). 
If the dependence on $\kappa$ could be, in principle, rationalized by reinterpreting activity as an effective temperature generating fluctuations that carry an energy that scales as $v_p^2$, and one would expect  $v_p^2>\kappa$, i.e $v_p>\kappa^{1/2}$, for the shell to begin to crumple, this simple argument does not, however, explain the weaker dependence on the stretching constant. 
Furthermore, despite our best attempts, we were unable to use temperature to crumple our shells in equilibrium. Although, partial buckling of the shell can be achieved by increasing the temperature of the bath at equilibrium~\cite{paulose2012,kovsmrlj2017}, our simulations at high temperatures show no crumpling of the shell. 

As a possible estimate of the size of the active fluctuations to be mapped into an effective temperature in our system, we considered a recent work on ideal active membranes which used the same model for an open elastic sheet. In that case an effective temperature was obtained by matching temperature and activity at the crumpling transition point, yielding an effective temperature  $T^{\text{eff}}=(1+1/42\,v_p^2)\,T_0$~\cite{gandikota2023}. For a bending of $\kappa=10$, the crumpling of active shells at $v_p^*\approx 30$ would then correspond to $T^{\text{eff}}\approx 22\,T_0$. Alternatively, one could try to estimate the extent of the active fluctuations using the active energy scale $k_{s}T_{s}=\gamma v_p^2/(6D_r)$~\cite{takatori_swim_2014}. In this case one would obtain an effective temperature $T^{\text{eff}}\approx 55\,T_0$.
We however find that even at temperatures of the thermal bath reaching values as large as $100-200\,T_0$, passive equilibrated shells do not show signs of crumpling. For instance, at $T=100\,T_0$ , even when we provide a crumpled shell as the initial configuration, the volume of the shell re-swells over time to 60\% of its original volume (see Appendix~\ref{shell_reswell}). This suggests that the crumpled conformation of the shell is distinctly non-equilibrium in nature, and a simple temperature mapping of the activity is not sufficient to explain our data. 

\section{The Flory phase}
\begin{figure}[t]
	\centering
	\includegraphics[width=0.4\textwidth]{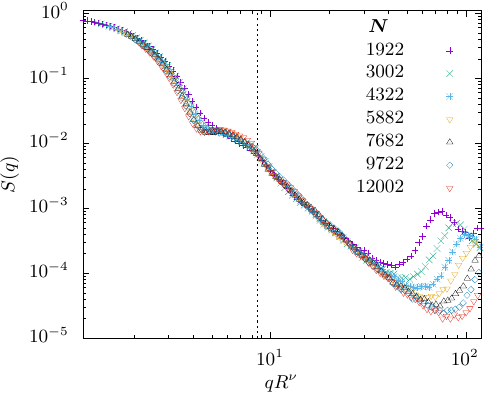}
	\caption{Log-log plot of the structure factor $S(q)$ for different shell sizes $N$, at $v_p=100$, as a function of the rescaled wave vector $qR^{\nu}$.
    The best collapse of the data for $q>2\pi/R_g$ is shown in this figure and is obtained  for $\nu=0.76\pm0.06$ which is within error bar of the Flory exponent $\nu=4/5$. The elastic constants are 
     $K=160$ and $\kappa=10$.}\label{struct_factor}
	\end{figure}
So far, we have used the term {\it crumpled} quite loosely and with  the only intent of providing a visual description of the shape of the shell at large active forces.
To accurately characterize the physical properties of this phase, we now calculate the size exponent, $\nu$, for the shells deep into the crumpled phase. For a shell with elastic constants $K=160$ and $\kappa=10$ discussed in Fig.~\ref{finite}, we set $v_p=100$.
We evaluate the size exponent by computing the shell structure factor $S(q)$ for different values of the shell radii $R\propto N^{1/2}$. We then plot $S(q)$  against $qR^{\nu}$, and find the value $\nu$ for which all data points collapse onto the same curve within the range $2\pi/R{\rm g}$ and $2\pi/\sigma$. The results of this analysis are shown in Fig.~\ref{struct_factor}. The best collapse is obtained for $\nu=0.76(6)$, a result within error bars of the Flory exponent $\nu=4/5$ predicted for the crumpled phase of self-avoiding membranes, and clearly different from $\nu=1$ and $\nu=2/3$ associated with the flat and the compact phase, respectively. 

\begin{figure}[t]
	\centering
	\includegraphics[width=0.38\textwidth]{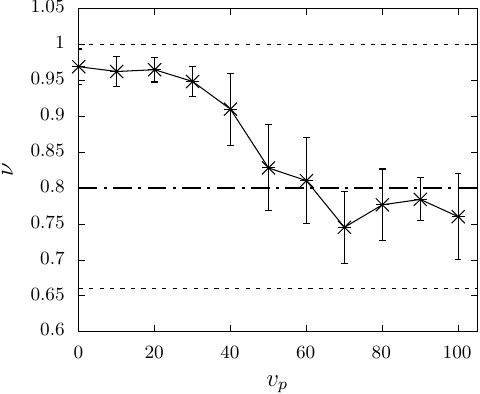}
	\caption{Size exponent of an elastic shell $\nu$ as a function of the self-propulsion speed $v_p$. 
     The elastic constants are $K=160$ and $\kappa=10$.  The dotted lines indicate the size exponents of a spherical ($\nu=1$) and a compact ($\nu=2/3$) phase. The dash-dotted line at $\nu=4/5$ marks the crumpled Flory phase. While the shell is locally flat for $v_p\leq v_p^*=30$, the size exponent for larger values of  $v_p$ show a clear departure from the flat phase and saturates to $\nu \approx 0.76$ for $v_p>v_p^*$.}\label{pe_nu}
	\end{figure}

We also evaluated how the size exponent of the shell depends on the strength of the active forces along the full spectrum of the shape transformation. 
What we find is that the size exponent crosses over  continuously from $\nu=1$ in the icosahedral/spherical state, to $\nu=0.76$ in the collapsed state. We do observe somewhat faster change beyond $v_p=30$ which we had identified as the onset self-propulsion speed for crumpling.
The results are shown in Fig.~\ref{pe_nu}.
Given the computational cost of these calculations, the values of $\nu$ in
Fig.~\ref{pe_nu} for the largest values of $v_p$ are obtained by collapsing the power-law regime of the structure factor as discussed above using five curves (rather than seven) associated with shells of size $N=1922,3002,4322,5882,7682$. We calculate the exponents and error bars using a measure that quantifies the quality of  collapse~\cite{bhattacharjee2001measure}. 
Given that for $v_p \leq 30$, the shell has a simpler icosahedral shape, the structure factor has multiple peaks and does not exhibit a power-law behaviour. In this low-activity regime, we instead find the size exponent  by fitting the size scaling of the radius of gyration $R_g(R)=aR^\nu$.

\section{Conclusions}
In this paper, we carried out extensive numerical simulations to understand the role of active fluctuations on the structural properties of thin elastic shells. Remarkably, we observe that thin shells easily and fully crumple under the presence of active fluctuations. Furthermore, we discovered that for different elastic constants, the curves describing the relative volume change of the shells as a function of the strength of the active forces, collapse into a single universal curve when appropriately normalized. Along this curve, the size exponent continuously decreases from $\nu=1.00$ to $\nu=0.76\pm0.06$, and the latter is compatible with the elusive crumpled Flory phase postulated for thermalized self-avoiding elastic membranes. While thermalized membranes are found to depart their flat phase only with explicit attractive interactions~\cite{abraham1991folding}, active shells which are distinctly out of equilibrium are seen to crumple even in the absence of explicit attractive interactions.

The crumpled structures of quasi non-extensile surfaces such as paper or aluminium-foil unattainable by simple equilibration of their microscopic counterparts can also only be formed by non-equilibrium means and are found to be in the Flory phase~\cite{gomes1987,kantor1987,gomes1989}. While these remain static in a quenched disordered state, the crumpled phase of active shells continuously fluctuates exploring the ensemble of crumpled configurations.

It is known that sufficiently large thin spherical shells buckle due to an effective negative pressure generated by thermal fluctuations~\cite{paulose2012,kovsmrlj2017}. However, in the absence of an explicit internal pressure~\cite{vliegenthart2011compression}, we found no evidence of crumpling in such equilibrated shells even when the strength of the thermal fluctuations become comparable to that of the harmonic bonds that keep the shell together. This is not surprising as a crumpled phase at such large temperatures would be entropically unfavorable,  
suggesting that activity cannot be mapped into an effective temperature for thin spherical shells. 
For equilibrated shells with large negative internal pressure, the shells collapse into shapes with a size exponent of $\nu=1.00\pm0.03$ (see Appendix~\ref{press_coll}). Thus the crumpled phase of active shells cannot be qualitatively mapped to shells that are collapsed using explicit inward pressure.
The crumpling of active shells is a particularly intriguing result because we have recently shown that active elastic sheets (rather than spherical shells) behave similarly to high temperature passive sheets~\cite{gandikota2023}. Specifically, a crumpled phase was not found in that instance as the sheet remained extended for all self-propulsion velocities considered in that study. 
Our results thus raise questions about the role of intrinsic curvature and topology in these systems. More work is currently underway to sort out their roles.

It is worth noting that studies of the one-dimensional analogs of our system, active rings in two dimensions, present a re-entrant behavior of the radius of gyration with activity. In that case, a narrow region for intermediate activities where the ring collapses has also been observed, but this is immediately followed by a re-swelling  at larger activities~\cite{mousavi2019active}. We verified that re-swelling does not occur in our system by running a few simulations at larger activities $v_p=200, 400$ (data not shown), making the two-dimensional shell qualitatively different from its lower dimensional counterpart.

More recently, the role of non-equilibrium fluctuations on the shape of an elastic shell has also been tested by performing Monte Carlo simulations with an explicit detail-balance breaking rule~\cite{agrawal2023active}. In this case, buckling of the shell was observed as a function of the degree of detailed balance breaking, but a crumpled phase was not observed.

Potential realizations of active shells could be constructed with water permeable elastic capsules using polymers. In principle one could use porous cross-linked polymeric vesicles with tethers connecting active particles to their surface, or cross-polymerized colloidosomes built from active particles.

\section*{Acknowledgements}
A.C. acknowledges financial support from the National Science Foundation under Grant No. DMR-2003444. We acknowledge useful discussions with Andrej Ko\v smrlj.

\bibliography{References}
\bibliographystyle{apsrev4-1}

\renewcommand\thefigure{SI.\arabic{figure}} 
\setcounter{figure}{0}
\renewcommand\theequation{SI.\arabic{equation}} 
\setcounter{equation}{0}

\begin{figure*}[t]
	\centering
	\includegraphics[width=0.85\textwidth]{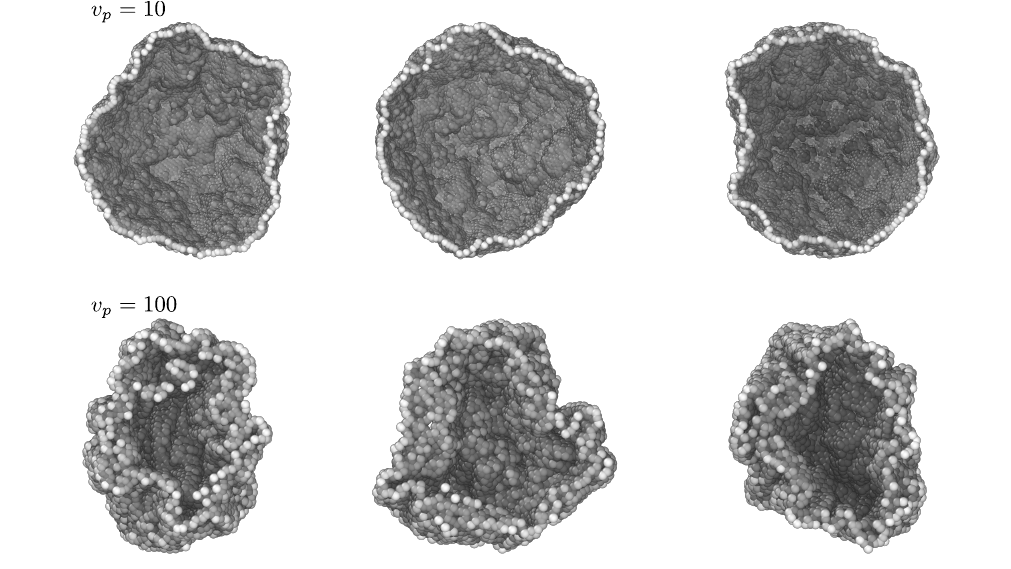}
	\caption{Cross-sections of active elastic shells at $v_p=10$ (top row) and $v_p=100$ (bottom row). The perspectives are along three orthogonal axis. The elastic constants are 
     $K=160\,k_{\rm B}T_0/\sigma^2$ and $\kappa=10\,k_{\rm B}T_0$ for an amorphous shell of size $N=12002$. }\label{cross-section}
 \end{figure*}

\newpage
\section{Appendix}

\subsection{Discretizing the Laplacian operator to implement bending energy}\label{discretization}
\begin{figure}[h]
	\centering
	\includegraphics[width=0.45\textwidth]{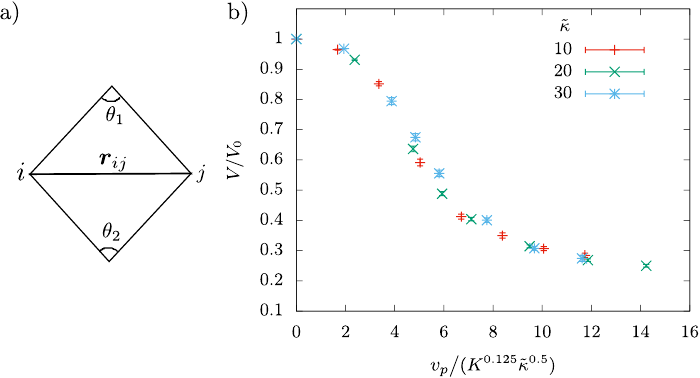}
	\caption{a) Sketch of two flat triangles sharing a bond. b) The elastic stretching constant $K=160$ and the size of the crystalline shell is $N=1922$. }\label{sigmaij}
 \end{figure}

When the shell is discretized using flat triangles, bending forces can be calculated using two methods - referred to as Method A and B~\cite{guckenberger2016bending,guckenberger2017theory}. For all numerical calculations, we implemented Method~A where we calculate the angle between normals of neighboring triangles (see Eq. 1 of the main text). To ensure that the collapse of normalized volumes of the shells as a function of $v_p/(K^{0.125}\kappa^{0.5})$ is not a numerical artefact of our choice of using Method A, in this section, we present some results where we calculate bending forces using Model~B~\cite{gompper1996random,guckenberger2017theory}. For a membrane whose local mean curvature at $\bm{r}$ is $H(\bm{r})$, the height field is $h(\bm{r})$ and the bending constant is $\tilde{\kappa}$, the Helfrich bending energy is, 
\begin{equation}
    E_{\text{bend}}= \frac{\tilde{\kappa}}{2} \int dA \,H(\bm{r})^2\approx\frac{\tilde\kappa}{2}\int dA \,[\nabla^2 h(\bm{r})]^2.
\end{equation}
 For a triangulated surface, whose node positions are denoted by $\bm{r}_i$, the discretized form of the Laplacian allows the mean curvature at the $i^{\text{th}}$ vertex to be calculated as,
\begin{equation}
    H_i=\frac{1}{\sigma_i} \bm{n}_i\cdot\sum_j \sigma_{ij}\frac{\bm{r}_{ij}}{r_{ij}},
\end{equation}
where index $j$ denotes the neighboring vertices of $i$,  $\bm{r}_{ij}=\bm{r}_i-\bm{r}_j$, $r_{ij}=|\bm{r}_{ij}|$, $\bm{n}_i$ is the surface normal and $\sigma_i$ is the area of the virtual dual cell of vertex $i$,
\begin{equation}
    \sigma_i=\frac{1}{4}\sum_j \sigma_{ij}r_{ij}.
\end{equation}
The length $\sigma_{ij}$ is, 
\begin{equation}\label{theta}
    \sigma_{ij}=\frac{r_{ij}}{2}[\text{cot}(\theta_1)+\text{cot}(\theta_2)],
\end{equation}
where $\theta_1$ and $\theta_2$ are opposing angles of the two triangles sharing the bond $ij$. See Fig. \ref{sigmaij}(a).
The discretized bending energy we use in our simulations is,

\begin{equation}
    E_{\text{bend}}=\frac{\tilde{\kappa}}{2}\sum_i \frac{1}{\sigma_i}\left[\bm{n}_i\cdot\sum_j \sigma_{ij}\frac{\bm{r}_{ij}}{r_{ij}}\right]^2. 
\end{equation}
In Fig. \ref{sigmaij}(b), we see that the collapse of the normalized volumes as a function of $v_p/(K^{0.125}\tilde{\kappa}^{0.5})$ is retained despite the use of an alternate implementation of bending energy.

\subsection{Crumpled active shells reswell on removal of activity}\label{shell_reswell}
\begin{figure}[h]
	\centering
	\includegraphics[width=0.45\textwidth]{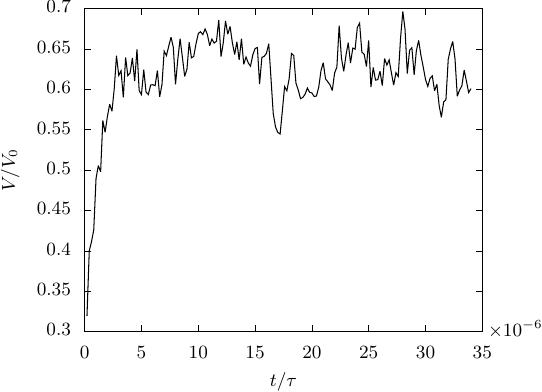}
	\caption{Normalized volume $V/V_0$ as a function of time depicting the reswelling of a crumpled configuration. The elastic constants are 
     $K=160\,k_{\rm B}T_0/\sigma^2$ and $\kappa=10\,k_{\rm B}T_0$ and the size of the amorphous shell is $N=1922$.}\label{reswell}
\end{figure}

Elastic shells buckle at high temperatures of $T=100\,T_0$, yet do not display a crumpled phase as we verified. To check the stability of the crumpled phase at high temperatures, we use a crumpled configuration as the initial configuration for a shell in a thermal bath of high temperature. Specifically, the initial configuration was generated by subjecting the shell to active fluctuations of strength $v_p=100$ at $T=\;T_0$ after which the activity was removed and the temperature is increased to $T=100\,T_0$. The shell immediately reswells to a volume of $V\approx0.6\, V_0$. See Fig.~\ref{reswell}. This suggests that a crumpled phase of the elastic shell is not stable even under these high temperatures.

\subsection{Finite size dependence}
\subsubsection{Radius of gyration}
\begin{figure}[h]
	\centering
	\includegraphics[width=0.45\textwidth]{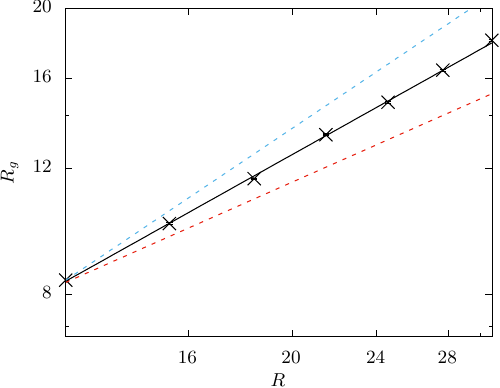}
	\caption{At a self-propulsion speed of $v_p=100$, the radius of gyration of active crystalline shells scales with shell radii as $R_g\sim R^\nu$ with $\nu=0.83\pm0.03$. In this log-log plot, the dashed blue and red lines are for reference and show the size scaling of the flat phase and  compact phase respectively. The small error bars on the data are standard errors.
 }\label{rg_scaling}
\end{figure}

At every self-propulsion speed $v_p$, the radius of gyration $R_g$ must scale with radius of spherical shell $R$ as $R_g(v_p)\sim R^{\nu(v_p)}$ where the size exponent $\nu(v_p)$ is numerically calculated in Fig. 4 of the main text. For $v_p=100$, the size dependence of the radius of gyration is shown in Fig. \ref{rg_scaling} where the best fit for data is at $\nu(100)=0.83\pm0.03$. The error bar is the fitting error.

\subsubsection{Normalized volumes}\label{norm_vol}
If the volume $V(v_p)$ of the shell is approximated as $V(v_p)\sim R_g^3(v_p)$, then 

\begin{equation}\label{volume_scale}
    \frac{V(v_p)}{V_0} \approx \frac{R_g^3(v_p)}{R^3} = R^{3[\nu(v_p) -1]}.
\end{equation}

\noindent For small self propulsion speeds, $\nu(v_p)-1 \approx 0$ and following Eq. \ref{volume_scale}, the normalized volume curves should exactly overlap irrespective of size of the shell. See Fig. 1 of main text. 
With increasing $v_p$, $\nu(v_p)-1$ monotonically decreases as seen in Fig. 4 of main text. Thus, following Eq. \ref{volume_scale},  
the normalized volumes $V(v_p)/V_0$ become more size dependent with increasing $v_p$. Correspondingly, the spread of the normalized volume curves of different sizes is seen to increase with increasing $v_p$ as can be seen in Fig.~\ref{finite} of the main text.

\subsection{Pressure induced collapse of shells}\label{press_coll}
\begin{figure}[h]
	\centering
	\includegraphics[width=0.35\textwidth]{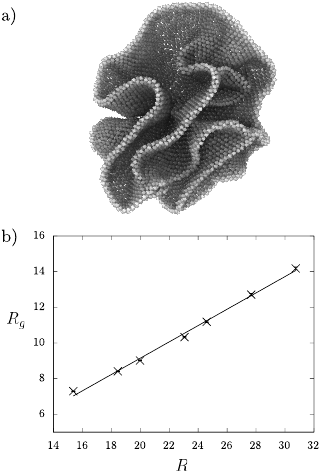}
	\caption{a) Snapshot of a pressure collapsed crystalline shell of size $N=12002$.  b) At a large inward pressure of $p/p_c=200$, the radius of gyration scales with the shell radii as $R_g\sim R^\nu$ with $\nu=1.00\pm 0.03$. The small error bars on the data are standard errors. Note that both axes in the plot are linear.
 }\label{pressure_collapse}
\end{figure}

Passive shells can be collapsed with sufficient negative i.e. inward pressures. To find qualitative differences between activity-induced collapse and pressure-induced collapse, we calculate the size exponent of shells that are collapsed using negative pressures. 
For this elastic shell under the influence of pressure $p$, we perform Monte Carlo simulations using the interaction potential, 
\begin{equation}
	\begin{split}
		U_p&=K_s\sum_{<ij>}\Theta(r_{ij}-l_0)+\kappa\sum_{<lm>}(1-\bm{\eta}_l\cdot\bm{\eta}_m) - pV\\
		   &+ 4\,\varepsilon\sum_{ij}\left[ \left( \frac{\sigma}{r_{ij}}\right)^{12} - \left(\frac{\sigma}{r_{ij}}\right)^{6} +\frac{1}{4}\right].
	\end{split}
\end{equation}
All variable definitions are as defined under Eq.~\ref{Hamiltonian} of the main text. The Heaviside step function $\Theta(x)=1$ for $x\geq 0$ and zero otherwise while the constant $K_s\rightarrow\infty$. 

To collapse the shell, we use a large pressure $p/p_c=200$ for all sizes, where the critical pressure for buckling scales with shell radius as $p_c\sim 1/R^2$~\cite{LandauBook}.  
With the shell in a thermal bath of temperature $T=T_0$, we relax the shell until the radius of gyration saturates to a constant value after which we calculate the average radius of gyration $R_g$. See Fig. \ref{pressure_collapse}(a) shows
the typical shape of highly compressed elastic shells.
The scaling of the radius of gyration with shell radii is $R_g\sim R^\nu$ with $\nu=1.00\pm0.03$. The error bar is the fitting error. See Fig. \ref{pressure_collapse}(b) where we show the averaged data for twenty independent runs. The shell despite its complete collapse scales linearly with shell radii in contrast to active shells that remain in the crumpled phase at high self-propulsion speeds.

 \end{document}